\documentclass[12pt]{JHEP}
     \usepackage{amssymb} 
     \usepackage{amsmath}
     \usepackage{graphicx}
     \usepackage{epsf}
\catcode `@=11 \@addtoreset{equation}{section} \catcode `@=12

     \def\IP{\relax{\rm I\kern-.18em P}}

     \def\nn{\nonumber}                  

     \def\cedille#1{\setbox0=\hbox{#1}\ifdim\ht0=1ex \accent'30 #1%
      \else{\ooalign{\hidewidth\char'30\hidewidth\crcr\unbox0}}\fi}
     
      \def\ad{\mbox{\rm ad}}

     \def\H{{\cal H}}

\def\be{\begin{equation}}
\def\ee{\end{equation}}
\def\ba{\begin{array}{l}}
\def\ea{\end{array}}
\def\bea{\begin{eqnarray}}
\def\eea{\end{eqnarray}}
\def\beas{\begin{eqnarray*}}
\def\eeas{\end{eqnarray*}}
\def\eq#1{(\ref{#1})}

\def\nn{\nonumber\\}

\def\ket#1{| #1 \rangle}

\def\gap#1{\vspace{#1 ex}}

\def\half{\frac{1}{2}}

\def\psid{\psi^\dagger}

\def\ad{a^\dagger}
\def\sigmad{\sigma^\dagger}
\def\betad{\beta^\dagger}

\def\dcff#1#2{D(#1| #2)}
\def\gcff#1#2{\Gamma(#1| #2)}
\def\tgcff#1#2{\tilde \Gamma(#1| #2)}
\def\ads{$AdS_5 \times S^5$}

\title{From Gravitons to Giants}
\author{Avinash Dhar, Gautam Mandal and Mikael Smedb\"ack \\  
Tata Institute of Fundamental Research, Homi Bhabha Road, \\
Mumbai 400 005, India 
\\~~\\
\email{adhar@theory.tifr.res.in, mandal@theory.tifr.res.in, smedback@theory.tifr.res.in
}}

\preprint{\hepth{0512312}\\
TIFR/TH/05-49
}
     
\abstract
{We discuss exact quantization of gravitational fluctuations in the
half-BPS sector around AdS$_5 \times $S$^5$ background, using the dual
super Yang-Mills theory. For this purpose we employ the recently
developed techniques for exact bosonization of a finite number $N$ of
fermions in terms of $N$ bosonic oscillators. An exact computation of
the three-point correlation function of gravitons for finite $N$ shows
that they become strongly coupled at sufficiently high energies, with
an interaction that grows exponentially in $N$. We show that even at
such high energies a description of the bulk physics in terms of
weakly interacting particles can be constructed. The single particle
states providing such a description are created by our bosonic
oscillators or equivalently these are the multi-graviton states
corresponding to the so-called Schur polynomials. Both represent
single giant graviton states in the bulk. Multi-particle states
corresponding to multi-giant gravitons are, however, different, since
interactions among our bosons vanish identically, while the Schur
polynomials are weakly interacting at high enough energies.}

\keywords{Bosonization, AdS-CFT, String Theory, Supergravity}

\begin{document}  
    

\section{Introduction}

In a recent work \cite{DMS} techniques have been developed for exact
bosonization of a finite number $N$ of nonrelativistic fermions.  This
has opened up new possibilities for exploring sectors of string theory
non-perturbatively.  One of the key motivations for this work was the
study by Lin, Lunin and Maldacena \cite{LLM} of a class of half-BPS
type IIB geometries in asymptotically AdS spaces\footnote{ Various
extensions of \cite{LLM} have been made. $1/4$-BPS excitations were
considered in \cite{quarter}, $1/8$-BPS excitations in
\cite{Berenstein} while bubbling geometries in $AdS_3$ were
investigated by \cite{ads3,D1D5}. $1/2$-BPS excitations of $AdS_5
\times \mathbb{R}P^5$ were considered in \cite{orientifolds}. Extension to
multi-charge $1/2$-BPS case has been considered in \cite{Yoneya}.}
and their connection with the semiclassical states of a free fermi
system. Taken together with \cite{DMS}, the LLM work offers an
excellent laboratory to explore non-perturbative aspects of quantum
gravity in the above sector. Such a study was initiated
\footnote{Related aspects of this issue were discussed in
\cite{babel-1,babel-2}. See also \cite{Silva,Shepard}.}
in \cite{DMS} using the exact bosonization methods developed there
\footnote{This bosonization works for arbitrary fermion Hamiltonian
and can also be applied to $c=1$, free fermions on a circle 
(the Tomonaga problem), etc. Also see the remarks at the end of Sec 2.}. In
the present work we will explore these issues in greater detail.

In the boundary super Yang-Mills theory, the states corresponding to
half-BPS LLM geometries are described by $N$ free fermions in a
harmonic potential
\cite{Corley:2001zk,Berenstein:2004kk,Takayama:2005yq} (see also \cite{KW}).
At large $N$,
there is a semiclassical description of the states of this system in
terms of droplets of fermi fluid in phase space; LLM showed that there
is a similar structure in the classical geometries in the bulk. This
semiclassical correspondence is already remarkable in the sense that
it exhibits a noncommutative structure of two of the space coordinates
\cite{DMS,LLM,Mandal:2005wv}; however, finite $N$ effects,
corresponding to fully quantum mechanical aspects of bulk gravity,
open up more interesting questions
\cite{DMS,Dhar:2005qh}. Semiclassically, small fluctuations of the
droplet boundaries correspond to small gravitational fluctuations
around the classical geometries, as anticipated by \cite{LLM} and
shown by \cite{Grant:2005qc,Maoz:2005nk} (in part using the symplectic
form calculated by \cite{poly2}).  At finite $N$ only those
fluctuations of the fermi system which have low enough excitation
energy compared to $N$ can be identified with gravity modes
propagating in the bulk. From the bulk gravity point of view, the
relevant (dimensionless) scale is $R/l_p \sim N^{1/4}$, where $R$ is
the AdS radius and $l_p$ is the ten-dimensional Planck scale, since
beyond these energies perturbative corrections may be expected to
become large. However, from an exact calculation in the boundary
theory we find that perturbation theory actually breaks down much
later, at energies of order $N^{1/2}$. It is possible that the reason
for this is cancellations due to the high degree of supersymmetry in
the half-BPS sector. A different argument for the existence of an
energy scale of order $N^{1/2}$ exists \cite{Bala} that suggests
breakdown of weakly coupled gravity picture for the LLM gravitons at
this scale. Essentially the argument is that the size of the
wavefunction (in AdS$_5$) of an LLM graviton excitation decreases with
energy and at an energy scale of order $N^{1/2}$ it becomes order
Planck scale.

At still higher energies, one would expect a description of the bulk
physics in terms of graviton excitations to break down. In fact, as we
will see here, at sufficiently high energies gravitons cease to make
sense as weakly coupled degrees of freedom since their correlations
grow exponentially with $N$. This happens long before graviton
energies are of order $N$. At sufficiently high energies, therefore,
we need to seek out new weakly coupled degrees of freedom which can
provide a more meaningful description of bulk physics than
gravitons. We will explicitly find these degrees of freedom in this
paper. As we will argue in this paper, there is strong evidence that
these new degrees of freedom are giant gravitons
\cite{McGreevy:2000cw,Grisaru:2000zn,Hashimoto:2000zp}. In the
boundary theory, the corresponding single-particle states are created
by the oscillators of the bosonized theory. Since these are strictly
non-interacting, one can obviously describe physics in terms of these
at all energies. Remarkably, we find that the single-particle states
created by the oscillators are also exactly the single-particle states
corresponding to the combinations of multi-graviton states (for
totally antisymmetric representaions) known as Schur polynomials. This
does not hold for multi-particle states. In fact, there are small but
non-zero correlations among the Schur polynomials at high energies,
which distinguishes them from the oscillator degrees of freedom. In
any case, giant gravitons, which are closely related to both these
boundary degrees of freedom, have the right properties to provide a
good description of the bulk physics at high energies. This transition
from low-energy graviton degrees of freedom to more microscopic
degrees of freedom at high energies is expected to happen in any
consistent theory of gravity. The remarkable thing about the LLM
system is that it provides us with a laboratory in which we can
actually see this happening in a very controlled and explicit fashion.

The organization of this paper is as follows. In Sec 2 we will
summarize the work of \cite{DMS} \footnote{The work in this paper
discusses two different exact bosonizations of the fermi system; here
we will limit our discussion to bosonization of the first type.} on
the exact bosonization of a finite number $N$ of nonrelativistic
fermions.  The presentation here is somewhat different and simpler.
When applied to the LLM sector, we find that the bosonized theory is
described by a free hamiltonian for $N$ bosonic oscillators. In Sec 3
we will discuss correlation functions of the modes of fermion spatial
density, which correspond to single trace operators in the boundary
super Yang-Mills theory.  These are the modes for small fluctuations
of the bulk metric around AdS$_5 \times$ S$^5$ which preserve the
half-BPS condition. We argue that the effective low-energy physics of
these modes is described by a field theory with a small cubic coupling
of O($1/N$). However, the large-$N$ perturbation theory breaks down
when graviton energies are of order $\sqrt N$. This is shown by doing
an exact calculation of graviton three-point function in Sec 3. This
calculation also shows that gravitons cease to provide a meaningful
description of the bulk physics at much higher energies, which may
still be only a small fraction of $N$. Instead, at these energies we
must use the giant gravitons to provide a meaningful description of
the bulk physics. In Sec
\ref{Schur} we show that the single-particle states created by our
bosonic oscillators are identical to the single-particle states
created by the Schur polynomials for totally antisymmetric
representaions. This is done by establishing a general relation
between multitrace operators and the bosonic oscillators acting on the
fermi vacuum. We end with a summary and some comments in Sec
\ref{discussion}.

\section{Review of exact bosonization}

In this section we will review the techniques developed in \cite{DMS}
for an exact operator bosonization of a finite number of
nonrelativistic fermions. The discussion here is somewhat different
from that in \cite{DMS}. Here, we will derive the first bosonization
of \cite{DMS} using somewhat simpler arguments, considerably
simplifying the presentation and the formulae in the
process. Moreover, the present derivation of bosonization rules is
more intuitive, making its applications technically easier.

Consider a system of $N$ fermions each of which can occupy
a state in an infinite-dimensional Hilbert space  $\H_f$.
Suppose there is a countable basis of $\H_f:
\{ \ket{m}, m=0,1, \cdots, \infty\}$. For example, this could
be the eigenbasis of a single-particle hamiltonian, $\hat h \ket{m} =
{\cal E}(m) \ket{m}$, although other choices of basis would do equally
well, as long as it is a countable basis. In the second quantized
notation we introduce creation (annihilation) operators
$\psi^\dagger_m$ ($\psi_m$) which create (destroy) particles in the
state $\ket{m}$. These satisfy the anticommutation relations
\be
\{ \psi_m, \psid_n\}= \delta_{mn}
\label{anticom}
\ee
The $N$-fermion states are  given by (linear combinations of)
\be
\ket{f_1, \cdots, f_N}= \psid_{f_1}\psid_{f_2} \cdots
\psid_{f_N} \ket{0}_F
\label{fermi-state}
\ee
where $f_m$ are arbitrary integers satisfying 
$0\le f_1 < f_2 < \cdots < f_N$, and 
$\ket{0}_F$ is the usual 
Fock vacuum annihilated by $\psi_m, m=0,1, \cdots, \infty$.

It is clear that one can span the entire space of $N$-fermion states,
starting from a given state $\ket{f_1, \cdots, f_N}$, by repeated
application of the fermion bilinear operators
\be
\Phi_{mn} = \psid_m \, \psi_n
\label{phi-mn}
\ee
However, the problem with these bosonic operators is that they are not
independent; this is reflected in the W$_\infty$ algebra that they
satisfy,
\be
[\Phi_{mn}, \Phi_{m'n'}] = \delta_{m'n}\Phi_{mn'} - \delta_{mn'}\Phi_{m'n}.
\label{winf}
\ee
This is the operator version of the noncommutative constraint $u*u=u$
that the Wigner distribution $u$ satisfies in the exact path-integral
bosonization carried out in \cite{DMW-nonrel}.

\begin{figure}[htb]
       \centering
       \includegraphics[height=7cm]{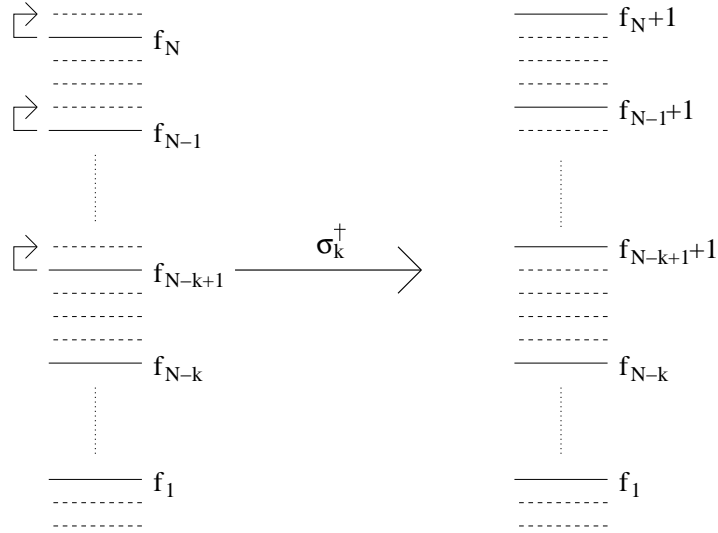}
       \caption{The action of $\sigma_{k}^{\dagger}$.}
       \label{fig:sigma}
     \end{figure}

A new set of unconstrained bosonic operators was introduced in
\cite{DMS}, $N$ of them for $N$ fermions. In effect, this set of
bosonic operators provides the independent degrees of freedom in terms
of which the above constraint is solved. Let us denote these operators
by $\sigma_k,~k=1, 2, \cdots, N$ and their conjugates,
$\sigmad_k,~k=1, 2, \cdots, N$. As we shall see shortly, these
operators will turn out to be identical to the $\sigma_k$'s used in
\cite{DMS}. The action of $\sigmad_k$ on a given fermion state
$\ket{f_1, \cdots, f_N}$ is rather simple.
It just takes each of the fermions in the top $k$ occupied levels 
up by one step, as illustrated in Fig.\ref{fig:sigma}.
One starts
from the fermion in the topmost occupied level, $f_N$, and moves it up
by one step to $(f_N+1)$, then the one below it up by one step, etc
proceeding in this order, all the way down to the $k$th fermion from
top, which is occupying the level $f_{N-k+1}$ and is taken to the
level $(f_{N-k+1}+1)$. For the conjugate operation, $\sigma_k$, one
takes fermions in the top occupied $k$ levels down by one step,
reversing the order of the moves. Thus, one starts by moving the
fermion at the level $f_{N-k+1}$ to the next level below at
$(f_{N-k+1}-1)$, and so on. Clearly, in this case the answer is nonzero
only if the $(k+1)$th fermion from the top is not occupying the level
immediately below the $k$th fermion , i.e.  if $(f_{N-k+1}-f_{N-k}-1) >
0$. If $k=N$ this condition must be replaced by $f_1 > 0$.

These operations are necessary and sufficient to move to any desired
fermion state starting from a given state. This can be argued as
follows. First, consider the following operator,
$\sigma_{k-1}~\sigmad_k$.  Acting on an arbitrary fermion state, the
first factor takes top $k$ fermions up by one level; this is followed
by bringing the top $(k-1)$ fermions down by one level. The net effect
is that only the $k$th fermion from top is moved up by one level. In
other words,
$\sigma_{k-1}~\sigmad_k=\psid_{f_{N-k+1}+1}~\psi_{f_{N-k+1}}=
\Phi_{{f_{N-k+1}+1},f_{N-k+1}}$. In this way, by composing together
different $\sigma_k$ operations we can move individual fermions
around. Clearly, all the $N$ $\sigma_k$ operations are necessary in
order to move each of the $N$ fermions indvidually. It is easy to see
that by applying sufficient number of such fermion bilinears one can
move to any desired fermion state starting from a given state.

It follows from the definition of $\sigmad_k,~\sigma_k$ operators that 
they satisfy the following relations:
\bea
\sigma_k~\sigmad_k= 1,~~\sigmad_k~\sigma_k=\theta(r_k-1),~~
[\sigma_l, \sigmad_k]=0,~~l \neq k,
\label{sigmas}
\eea
where $(f_{N-k+1}-f_{N-k}-1) \equiv r_k$ and $\theta(m)=1$ if $m \geq
0$, otherwise it vanishes. Moreover, all the $\sigma_k$'s annihilate the
Fermi vacuum.

Consider now a system of bosons each of which can occupy a state in an
$N$-dimensional Hilbert space $\H_N$. Suppose we choose a basis
$\{\ket{k},~k=1,\cdots,N\}$ of $\H_N$. In the second quantized
notation we introduce creation (annihilation) operators $a^\dagger_k$
($a_k$) which create (destroy) particles in the state $\ket{k}$. These
satisfy the commutation relations
\be
[a_k, \ad_l]= \delta_{kl}, \quad k,l=1, \cdots, N
\label{oscillator}
\ee 
A state of this bosonic system is given by (a linear combination
of)
\be
\ket{r_1, \cdots, r_N}= \frac{(a_1^\dagger)^{r_1}\cdots
(a_N^\dagger)^{r_N}}{ \sqrt{r_1 ! \cdots r_N!}} \ket{0}
\label{bose-state}
\ee

It can be easily verified that equations (\ref{sigmas}) are 
satisfied if we make the following identifications
\bea
\sigma_k &=& \frac{1}{\sqrt{\ad_k a_k + 1}} a_k, \nn
\sigmad_k &=& \ad_k\frac{1}{\sqrt{\ad_k a_k + 1}},
\label{sigmadefs}
\eea
together with the map
\bea
r_k &=& f_{N-k+1} - f_{N-k} - 1, \quad \quad k = 1,\,2,\,\cdots\,N-1 \nn
r_N &=& f_1.
\label{statemap} 
\eea
This identification is consistent with the Fermi vacuum being the
ground state of the bosonic system. The map \eq{statemap} first
appeared in \cite{Nemani}. The first of these arises from the
identification (\ref{sigmadefs}) of $\sigma_k$'s in terms of the
oscillator modes, while the second follows from the fact that
$\sigma_N$ annihilates any state in which $f_1$ vanishes.

Using the above bosonization formulae, any fermion bilinear operator
can be expressed in terms of the bosons. For example, the hamiltonian
can be rewritten as follows. Let ${\cal E}(m),~m=0, 1, 2, \cdots$ be
the exact single-particle spectrum of the fermions (assumed
noninteracting). Then, the hamiltonian is given by 
\be 
H = \sum_{m=0}^\infty {\cal E}(m) ~ \psid_m~\psi_m.
\label{fermiham}
\ee
Its eigenvalues are $E=\sum_{k=1}^{N} {\cal E}(f_k)$. 
Using $f_k=\sum_{i=N-k+1}^{N} r_i +k-1$, which is easily derived from 
(\ref{statemap}), these can be rewritten in terms of the bosonic occupation
numbers, $E=\sum_{k=1}^{N} {\cal E}(\sum_{i=N-k+1}^{N} r_i +k-1)$. These 
are the eigenvalues of the bosonic hamiltonian
\be
H = \sum_{k=1}^{N} {\cal E}(\sum_{i=N-k+1}^{N} \ad_i a_i +k-1).
\label{boseham}
\ee
This bosonic hamiltonian is, of course, completely equivalent to the 
fermionic hamiltonian we started with.

For the harmonic potential, the spectrum is linear. In this case we get
\be
H-H_F = \sum_{k=1}^{N} k \ad_k a_k,
\label{LLMham}
\ee
where $H_F$ is the Fermi ground state energy. This hamiltonian, and the
LLM system that it describes, will be the focus of our discussions in the
rest of this paper.

We remark that our bosonization technique does not depend on any
specific choice of fermion hamiltonian and can be applied to various
problems like $c=1$, free fermions on a circle (the Tomonaga problem)
\cite{DM} etc. Also see \cite{DMS} for some more details on this issue.

\section{Graviton interaction}

In this section we will present a quantum computation of graviton
correlators from the boundary theory.  The main result of this
computation, described in subsection \ref{breakdown}, will be to show
that for sufficiently high energy modes, the concept of gravitons
breaks down since the strength of their interaction grows
exponentially with $N$. We will show that such pathological behaviour
can be understood as a wrong choice of variables to describe gravity
at short wavelengths and the right variables to describe gravity at
such energies are the giant gravitons in terms of which the
interactions become weak.  Another result of the computation will be
to exhibit a chiral ring structure of the graviton interactions at low
energies, which reduces multipoint graviton interactions to
essentially a combination of the 'structure constants' of the chiral
ring.

\subsection{The exact correlators}

We recall that the standard AdS/CFT dictionary
identifies gravitons in the bulk to the single trace operators 
$Tr Z^m$ which are represented in the fermion theory
by the operators\cite{Takayama:2005yq}:
\bea
\beta_m^{\dagger} &&= 
\sum_{n=0}^{\infty} C(m,n) \psi_{n+m}^{\dagger} \psi_n \mbox{, }
\\
C(m,n)&& \equiv \sqrt{\frac{(m+n)!}{2^m n!}}
\label{def-beta}
\eea 
We will denote correlators of the theory as 
\be
\dcff{m_1,~m_2,...,~m_r}{n_1,~n_2,...,~n_s}
\equiv
\langle F_0 | \beta_{m_1}...\beta_{m_r} \betad_{n_1}...\betad_{n_s}
| F_0 \rangle
\equiv \langle \beta_{m_1}...\beta_{m_r} \betad_{n_1}...\betad_{n_s} 
\rangle
\label{dcffs}
\ee 
Here $| F_0 \rangle$ is the Fermi vacuum. An exact calculation of $\dcff{m,n}{m+n}$ 
and $\dcff{m}{m}$, valid for finite $N$, can
be done using either the fermion representation for the $\beta$'s
(which is simpler) or their bosonic representation in terms of the
oscillators $a_k,~\ad_k$ (see Appendix \ref{a.sec.betas}). 
We quote the results below \footnote{Our results agree with calculations 
done earlier using matrix model \cite{Kristjansen}.}:
\bea
\dcff{m,n}{m+n}&=&\frac{1}{2^{m+n}(m+n+1)} \biggl[ \frac{(N+m+n)!}
{(N-1)!}+
\frac{N!}{(N-m-n-1)!}
\biggr.
\nn
\biggl. &-&
\frac{(N+m)!}{(N-n-1)!}-\frac{(N+n)!}{(N-m-1)!}
\biggr],
\label{cmn1}
\eea
which is valid for $(m+n) < N$. For $(m+n)=N$, we get
\bea
\dcff{m,N-m}{N}=\frac{1}{2^N(N+1)} \biggl[ \frac{(2N)!}{(N-1)!}
-\frac{(N+m)!}{(m-1)!}-\frac{(2N-m)!}{(N-m-1)!} \biggr].
\label{cmn2}
\eea
Also,
\bea
\dcff m m=\frac{1}{2^m(m+1)} \biggl[ \frac{(N+m)!}{(N-1)!}-
\frac{N!}{(N-m-1)!} \biggr],~~m < N,
\label{zm1}
\eea 
and 
\bea
\dcff{N}{N}=\frac{1}{2^N(N+1)}\frac{(2N)!}{(N-1)!}.
\label{zm2}
\eea

The normalized correlators are given by
\be
\gcff{m_1,...,m_r}{n_1,...,n_s} = 
\frac{\dcff{m_1,...,m_r}{n_1,...,n_s}}
{\sqrt{\dcff{m_1}{m_1}..\dcff{m_r}{m_r}\dcff{n_1}{n_1}..\dcff{n_s}{n_s}
}}
\label{gammas}
\ee
We will mostly deal with $s=1$, that is, correlators involving only
several $Tr Z^{m_i}$ and one anti-holomorphic operator $Tr {\bar
Z}^n$. These satisfy a non-renormalization theorem and one
can consistently discuss them staying within the LLM sector
(loop corrections, which typically involve intermediate states outside
the LLM sector, are absent for these correlators).
For $s=1$, we have $n_s = n_1 =m_1 + ... + m_r$. 

\subsection{Graviton interaction at long wavelengths}

Let us first consider interaction of gravitons at low energies, where
the frequencies of the gravitons, $m,n$, will be considered to be
small numbers held fixed in the large $N$ limit.

We will begin by discussing the normalized correlators
\eq{gammas}, and in particular the 3-point function for the first
few modes $m, n=1,2,3...$. Some examples, 
obtained using the exact formulae \eq{cmn1},\eq{zm1} are
\bea
\dcff11 &&= N/2,~\dcff22 = N^2/2,~\dcff33 = 3(N^3+N)/8
\nn
\gcff{1,1}2 &&= \frac{\sqrt 2}N
\label{d-values}
\eea
The general behaviour of the 3-point functions for various
ranges of $m,n$ is discussed in Section \ref{breakdown}.
The fall-off with $N$ of the three-point function in \eq{d-values}, namely
$1/N$, is an example of the more general $\sqrt{mn(m+n)}/N$ behaviour
mentioned there.

We also need the values of some normalized four-point functions, 
which  can be calculated either 
using the formulae \eq{def-beta} or the relation between
the $\beta$'s and the $a, \ad$ oscillators indicated in
Appendix \ref{a.sec.betas}. A simple example is
\bea
\gcff{1,1,1}3 &&= \frac{2 \sqrt 3}
{ N^2}(1 + 1/N^2)^{-1/2} 
\eea

\subsubsection{Chiral ring structure}

The three-point functions \eq{cmn1} 
imply following chiral ring relation:
\bea
\beta_n~ \beta_m 
&=& \hat C_{mn} \beta_{m+n}[1 + O(1/N)]
\nn
\hat C_{mn} &\propto& 1/N
\label{ring}
\eea  
The `structure constant' is given by 
\bea
\frac{\dcff{m,n}{m+n}}{\dcff{m+n}{m+n}} =\hat C_{mn}[1 + O(1/N)] 
\label{c-mn}
\eea
which follows from computing the 
correlator of both sides of \eq{ring} with
$\betad_{m+n}$.
It is easy to see (using the approximation methods of Sec
\ref{breakdown}) that for large $N$ and fixed $m,n$, the 
leading behaviour of $\hat C_{mn}$ is $1/N$. We will provide
evidence for \eq{ring} by proving the following `bootstrap'
property of the 4-point function.

\subsubsection{Four-point function and `bootstrap'}

If the relation \eq{ring} is true, it follows that
the four-point function must satisfy what we will call a 
``bootstrap relation'' (see Fig.\ref{ope.fig} below):
\begin{figure}[ht]
\vspace{0.5cm}
\hspace{-0.5cm}
\centerline{
    \epsfxsize=10cm
   \epsfysize=4cm
   \epsffile{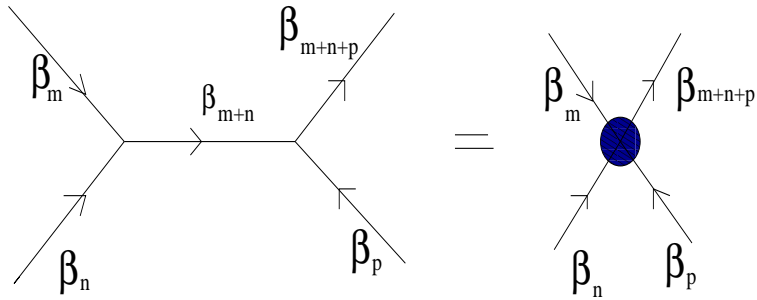}
 }
\caption{\sl Bootstrap:  Test for a cubic graviton field theory.}
\label{ope.fig}
\end{figure}

\bea
\langle \beta_m \beta_n \beta_p \betad_{m+n+p} \rangle
&& = \hat C_{mn} \langle \beta_{m+n} \beta_p \betad_{m+n+p} \rangle
[1 + O(1/N)]
\nn
&& = \frac{\dcff{m,n}{m+n} \dcff{m+n,p}{m+n+p}}{\dcff{m+n}{m+n}}
[1 + O(1/N)]
\nn
\label{bootstrap}
\eea

Example: For $m=n=p=1$,

\begin{equation}
  \begin{split}
    \text{LHS} & = \dcff{1,1,1}{3}= 3N/4, \\
    \text{RHS} & = \dcff{1,1}{2}\frac1{\dcff{2}{2}} \dcff{2,1}{3} =
    N/2\frac1{N^2/2}3N^2/4 = 3N/4. 
  \end{split}
\end{equation}
(In this case the
$O(1/N)$ correction is absent.)

Let us discuss  Eqn. \eq{bootstrap} in a little more detail.
On general grounds
\bea
\langle F_0 |\beta_m \beta_n \beta_p \betad_{m+n+p}| F_0 \rangle
=
\sum_n \langle F_0| \beta_m \beta_n |n\rangle \langle n|
\beta_p \betad_{m+n+p} | F_0 \rangle
\eea
where  $\sum_n |n\rangle \langle n|$ represents a sum
over all states
in the theory (we can restrict, for these
correlators, to states in the LLM sector,
which belong to the Fermion Fock space). 
Eqn. \eq{bootstrap} implies, to the leading
order in 1/N, that the only intermediate state that contributes
is $\betad_{m+n} \ket{F_0}$. This indeed turns out to be correct, 
because of the rather remarkable relation which is easy to prove,
\be
\beta_p \betad_{m+n+p}| F_0 \rangle= N^{p-1}\biggl[(1+O(\frac{1}{N}))\frac{p(p+m+n)}{2^p} 
\betad_{m+n}
| F_0 \rangle  + O(\frac{1}{N}) \betad \betad \cdots | F_0 \rangle \biggr]  
\label{remarkable}
\ee
where the last term is in general proportional to a
multi-graviton state (denoted by multiple $\betad$ acting on the vacuum).
To leading order in $1/N$, this term  is the 2-graviton state. 
We explicitly list a few examples: 
\bea
\beta_1 \betad_{3}| F_0 \rangle &&= \frac32 \betad_{2}
| F_0 \rangle 
\nn
\beta_1 \betad_{4}| F_0 \rangle &&= 2 \betad_{3}
| F_0 \rangle 
\nn
\beta_2 \betad_{4}| F_0 \rangle &&= N [2\betad_{2}
| F_0 \rangle + \frac{1}{N}(\betad_{1})^2 | F_0 \rangle ]. 
\eea

\subsubsection{Cubic gravity theory}

The above discussion about
the chiral ring suggests a field theory of gravitons
of the structure
\be
S = \int D_m \betad_m \beta_m + C_{mn} \betad_{m+n} 
\beta_m \beta_n  + c.c.
\label{cubic}
\ee
where the $D_m$ and $C_{mn}$ are related respectively to $\dcff{m}{m}$
and $\hat C_{mn}$, upto symmetry factors.  The notations $C_{mn}$,
$\hat C_{mn}$ and $C(m,n)$, defined respectively in (\ref{ring}),
(\ref{cubic}) and (\ref{def-beta}) are to be distinguished.

The cubic interactions in \eq{cubic} can be derived from  
\eq{exact} in $1/N$ expansion. They arise solely from the definition
of $\beta$'s as linear combinations of multi-$\ad$ oscillator states.
The theory \eq{cubic} has been matched to IIB supergravity
in \ads\ in the LLM sector in the work \cite{shiraz}.

A suggestion similiar to the above has also been made recently in the
work \cite{Okuyama}. Evidence in support of this suggestion has been
presented by matching direct computations of some correlation
functions in the fermion theory with cubic field theory.

\subsection{\label{breakdown}Breakdown of the graviton description}

In this section we will study the behaviour of the normalized
three-point function derived from \eq{cmn1}-
\eq{gammas}. In the discussion below we will restrict
ourselves to the case when all the three gravitons have energies 
much smaller than $N$. It turns out that there are three separate
energy regimes of interest. Let us consider the three cases in turn.

\begin{enumerate}
\item
$m, n$ fixed as $N \to \infty$. This is the large-$N$ perturbative regime.
Here
\bea
\gcff{m,~n}{m+n}=\frac{\sqrt{mn(m+n)}}{N}+O(1/N^2) 
\label{low-energy}
\eea
This result can be obtained by a straightforward $1/N$ expansion 
of \eq{cmn1}-\eq{zm2} and it agrees with
the calculation of tree-level three-graviton amplitude
in supergravity \cite{shiraz}.

\item 
$m/\sqrt N \sim n/\sqrt N$ fixed (and $O(1)$), as   $N \to \infty$. Here
\bea
\gcff{m,~n}{m+n} \approx a N^{-1/4}
\label{sqrtN-energy}
\eea
The quantity $a$ is a function of the fixed
ratio $m^2/N, a=f(m^2/N)$. This result has been obtained from 
the formula
\bea
\gcff{m,~n}{m+n} 
\approx \sqrt{\frac{m}N}\frac{\sinh^2\frac{m^2}N}{
\sinh\frac{m^2}{2N}\sqrt{\sinh\frac{2m^2}N}}, 
\label{3ptapprox}
\eea
which can be derived from \eq{cmn1}-\eq{zm2} using the 
formulae given in Appendix \ref{sec:3-point}. 
Although the overall strength of the interaction, $\sim
N^{-1/4}$, is small, to recover the function $f(m^2/N)$
(involving $\sinh(m^2/N)$ etc.) from
a perturbation theory in $1/N$, e.g. supergravity, one
needs to sum over all orders in $1/N$. We see that in this
regime of energies, perturbation theory breaks down 
(in the sense that no finite order calculation in $1/N$ will 
reproduce the above result).

\item
$m/N \sim n/N$ fixed and small as  $N \to \infty$. Here
\bea
\gcff{m,~n}{m+n} \approx a \exp[b N], ~a \propto \sqrt{m/N}, 
~b \propto m^2/N^2
\label{N-energy}
\eea
In this case the exponent $b$ turns out to be positive. Hence for $m$
a small fixed fraction of $N$, the correlator grows exponentially with
$N$! This means that gravitons become strongly interacting in this
energy regime.

\end{enumerate}

In fact, already for energies $m$ which grow as
$N^{1/2+\gamma}$, for $\gamma > 0$,  
the gravitons become strongly interacting 
and are not useful concepts as particles.
What, then, are the new weakly interacting entities
which can exist as perturbative states?

In the boundary theory that we are considering here, there are two
possible candidates. One is to replace gravitons by the oscillator
excitations of the bosonized theory discussed in Sec 2. These
excitations are exactly non-interacting and hence they can replace
gravitons at high energies. The corresponding bulk degrees of freedom,
at the level of single-particle states (see sec 3.4.1), are the giant
gravitons. The other possibility is to replace single-graviton states
by the Schur polynomial combinations of multi-graviton states (which
we will henceforth refer to merely as the ``Schurs''). The main reason for
this choice is that at high energies, Schurs are weakly
interacting. This can be seen from a calculation of multi-point
correlation functions of Schur poloynomials, $\chi_{m}(Z)$, which has
been done in \cite{Corley:2001zk}. From this work we can read off the
normalized 3-point function of the Schurs:
\be
\tgcff{m,~n}{m+n}
\equiv \frac{<\chi_{m}(Z)\chi_{n}(Z)\chi_{m+n}(\bar Z)>}{
||\chi_{m}(Z)||~||\chi_{n}(Z)||~||\chi_{m+n}(\bar Z)||}
=
\sqrt{\frac{ N!/(N-m -n)!}{N!/(N-m)!~ N!/(N-n)!}}
\ee
We see that
\begin{enumerate}
\item for $m, n$ fixed, as $N \to \infty$,
\be
\tgcff{m,~n}{m+n} \sim O(1)
\label{o-1-schur}
\ee
Clearly the Schur polynomials do not represent
weakly interacting particles for long wavelength 
modes. The gravitons (correspondig to single trace
operators), with interactions $\sim O(1/N)$,
are a better description of the low-energy perturbative spectrum.
However,

\item for $m/N \sim n/N$ fixed and small as $N \to \infty$, we get
\[
\tgcff{m,~n}{m+n} \sim e^{-a N}
\]
where $a$ is a positive quantity of $O(1)$. We see 
that Schurs have exponentially small interactions 
in this regime of energies, unlike the gravitons
whose interaction grows exponentially at such large energies.

For $m/N \sim 1$ fixed and $n=1$ \footnote{Here $n=1$ has
been taken for convenience. One could have taken it 
to be any number of order one, not necessarily
exactly one.}, the 3-point function of Schurs 
\footnote{It would be more appropriate to think of this
case as the coupling of a Schur to a graviton.}
goes as 
\[
\tgcff{m,1}{m+1}=\sqrt{1-m/N}.
\]
So even in this case there is a small but non-zero
interaction.  

\end{enumerate}

The above observations teach us two things: (i) description of
gravitons as perturbative spectrum breaks down for sufficiently large
energies (which are, however, still $\ll N$), (ii) there is,
nevertheless, a weakly coupled description available of the bulk
physics, not in terms of the old gravitons, but in terms of (the bulk
counterpart of) either the oscillator states of the bosonized theory
or the Schurs. It turns out that the Schurs are closely related to the
single-particle states created by the bosonic oscillators from the
fermi vacuum, a fact that we will prove in section
\ref{Schur}. Remarkably, therefore, both the choices lead to giant
gravitons as the right choice of degrees of freedom to replace
gravitons at high energies.

\subsection{The universal bosonic excitations}

It would appear from the above discussion of the
three-point functions that there is
a change of description of the perturbative spectrum
from gravitons to giant gravitons as one tunes 
the energy up from low to sufficiently high.
In principle, we could describe bulk physics 
at all energies in terms of bulk duals of the oscillators of the 
bosonized theory which create `particle' states whose
interaction strictly vanishes. In the boundary description
these are the particle states
created by the bosonic operators $\ad_{m_1}, \ad_{m_2}...$
These have an energy cut-off $m=N$ by construction. The hamiltonian
is exactly diagonal in terms of these oscillators
\bea
H && = \sum_{k=1}^{N} k \ad_k a_k
\nn
&& 
= \frac{1}{2} (N+1) \sum_{\mu=0}^{N-1} 
\phi^\dagger_\mu~\phi_\mu 
+ \frac{1}{2} \sum_{\mu \neq \nu=0}^{N-1} [1-i~\cot{\frac{\pi}{N}(\mu-\nu)}]~
\phi^\dagger_\mu~\phi_{\nu}
\label{exact}
\eea
The second equality above is the ``coordinate space'' representation of 
the same hamiltonian, where $\phi_\mu$ is a  ``lattice'' Fourier transform 
of $a_m$ \cite{DMS}.

This implies that there is a universal description of the perturbative
spectrum in the half-BPS sector in terms of states which
are non-interacting at all energies (with an in-built cut-off at
$m=N$). Both the single trace operators and Schur polynomials create
states which are linear combinations of these states.

\subsubsection{Bulk interpretation of the states $\ad_{m_1}
\ad_{m_2}...\ket 0$ \label{bulk-map}}

It is clear that the bulk map of the 
states $\ad_{m_1} \ad_{m_2}...\ket 0$ is a linear
combination of graviton states, as given by the equations 
\eq{singleparticle}-\eq{examples-many} in Appendix \ref{a.sec.betas}. Although the
$\ad$-particle states are free, the
gravitons interact because of such linear combinations. E.g.,
using \eq{examples} and \eq{examples-many}, we get 
\bea
\langle  F_0 | \beta_1 \beta_1 \betad_2 | F_0 \rangle
&& = 
 \langle 0 | (\half \sqrt{N(N-1)}\sigmad_2 + \half\sqrt{N(N+1)}
(\sigmad_1)^2) \times
\nn
&& (-\half \sqrt{N(N-1)} \sigmad_2
+ \half\sqrt{N(N+1)} (\sigmad_1)^2 )\ket 0 
\nn
&& = \frac{N}{2}
\eea 
Normalizing according to \eq{gammas}, we recover $\gcff{11}2$ 
as given in \eq{d-values}. Note that $\langle 0| (a_1)^2)\ad_2 | 0 \rangle
= 0$, as expected of strictly independent particle states.

As one considers higher and higher modes $m$, each $\beta_m$ involves
a larger number of  multi-$\ad$ states
and the graviton interactions  become stronger.

The states $\ad_{m_1} \ad_{m_2}...\ket 0$ have a closer relation to
the states created at the boundary by the Schur polynomials. Indeed,
as we will see in the next section \ref{Schur}, the single-particle
states $\ad_n \ket 0$ are identical to the states created by the Schur
polynomials for totally antisymmetric representations. 

Multiple Schurs create states which are not identical to multi-$\ad$
states (see, e.g. (\ref{multi-schur})).  Since the giant gravitons do
not appear to have perturbative open string excitations in the
half-BPS sector \cite{das}, it is likely that the giant gravitons do
not interact perturbatively.  This behaviour is consistent with the
multi-$\ad$ states at the boundary, since these are completely
non-interacting. Also, these boundary states have an inherent energy
cut-off, consistent with the giant gravitons. Unlike these states, the
Schur states are weakly interacting at high energies, while they
interact strongly at low energies. It would seem from these
considerations that it is the multi-$\ad$ states which corresponds to
multiple giant gravitons. However, this needs to be confirmed by
direct calculations of giant graviton interactions in the bulk
string/gravity theory.

 \subsection{\label{Schur}Schurs and the bosonic oscillators} 

In this section, we will discuss a relation between the single-particle
bosonic excitations and Schur polynomial excitations.
From our discussion in Sec 2, explicit formulae for the graviton operators
$\beta_m^{\dagger}$, Eqn. (\ref{def-beta}), acting on the fermi vacuum can
be easily translated into bosonized formulae in terms of the bosonic
oscillators $\sigma_i^{\dagger}$ acting on the vacuum state. Some examples
of this have been given in Appendix \ref{a.sec.betas}. These can
then be inverted to express the latter in terms of the former.

We can get an idea of the meaning of the $\sigma_k^{\dagger}\ket{0}$ states
by explicitly calculating them for a few small values of $k$ in terms of 
multiple $\beta$'s acting on the fermi vacuum. 
Taking appropriate linear combinations of
$\beta_1^{\dagger} \ket{F_0} $, $\beta_2^{\dagger} \ket{F_0} $,
$\beta_3^{\dagger} \ket{F_0} $, $\left( \beta_1^{\dagger} \right)^2
\ket{F_0} $, $\left( \beta_1^{\dagger} \right)^3 \ket{F_0} $ and
$\beta_1^{\dagger} \beta_2^{\dagger} \ket{F_0} $, we find that
\begin{equation}\begin{split}
  C(1, N-1) \sigma_1^{\dagger} \ket{0} & =
  \beta_1^{\dagger} \ket{F_0} \mbox{, }  \\ C(2, N-2) \sigma_2^{\dagger}
  \ket{0} & = \frac{1}{2!} \left[ \left( \beta_1^{\dagger} \right)^2 -
  \beta_2^{\dagger} \right] \ket{F_0} \mbox{, } \\ C(3,
  N-3)\sigma_3^{\dagger} \ket{0} & = \frac{1}{3!} \left[ \left(
  \beta_1^{\dagger} \right)^3 -3 \beta_1^{\dagger} \beta_2^{\dagger}
  +2 \beta_3^{\dagger} \right] \ket{F_0} \mbox{,}  
\label{eq:a1a2a3}
\end{split}
\end{equation}
where $C(m,n)$ are defined in (\ref{def-beta}).
These can be generated from the formula
\begin{equation}\label{eq:recursion}
  \sum_{m=1}^k (-1)^{k-m} \beta_m^{\dagger}
  \tilde{\sigma}_{k-m}^{\dagger} \ket{0} + (-1)^k k
  \tilde{\sigma}_k^{\dagger} \ket{0} = 0,
\end{equation}
(for $k=1,2,3$) where
\begin{equation}
  \tilde{\sigma}_k^{\dagger} \equiv C(k,N-k) \sigma_k^{\dagger}.
\end{equation}
Note that in writing the equations \eq{eq:a1a2a3} and
\eq{eq:recursion} we have implicitly used the fact that the Fermi
vacuum, $\ket{F_0}$, and the bosonic vacuum, $\ket{0}$, are the same
state. It thus makes sense to have the $\beta$'s acting on either
$\ket{F_0}$ or $\ket{0}$.

As is proven in Appendix \ref{sec:proofrec}, the recursion relation 
(\ref{eq:recursion}) 
is actually valid at a general level $k=1,2,\ldots$, 
and so generates all the $\tilde{\sigma}_k^{\dagger}
\ket{0} $'s in terms of multiple $\betad_k$'s acting on the vacuum.
This means that the single-particle states created
by the bosonic oscillators $\sigma_k^{\dagger}$ acting on the vacuum are
identical to the single-particle states created by the Schur
combinations of single and multi-particle graviton states.
This is because the operators on right-hand side of (\ref{eq:a1a2a3}) have precisely the
form of Schur polynomials in the completely antisymmetric representation. Denoting the Schurs
by $s_k$, where $k=1, 2, 3, \cdots$, we have
\begin{equation} \begin{split}
    \mbox s_1 & = \mbox{Tr}X \text{, } \\
    \mbox s_2 & = \frac{1}{2} \left[ \left( 
    \mbox{Tr}X \right)^2 - \mbox{Tr}X^2 \right] \text{, } \\
    \mbox s_3 & = \frac{1}{6} \left[ \left( \mbox{Tr}X \right)^3-3 
    \mbox{Tr}X^2 \mbox{Tr}X +2 \mbox{Tr}X^3\right] \text{, } 
\end{split}
\end{equation}
etc. where $X$ is a hermitian matrix. These relations follow from 
the recursion formula \cite{Candelas}
\begin{equation}
\sum_{m=1}^k (-1)^{k-m} \mbox{Tr} X^m s_{k-m} + (-1)^k k s_k 
= 0,
\label{schur-recur}
\end{equation}
which defines all the Schur polynomials 
in the completely antisymmetric
representation\footnote{ The Schur polynomials in the completely
antisymmetric representation are also known as Chern polynomials
\cite{Candelas,FH}. In \cite{Candelas}, the latter are denoted by  
$c_k$, and appear in the
context of topological classifications of manifolds.  The matrix under
consideration is formed from components of the Ricci 2-form
$\mathcal{R}_{\alpha \beta \gamma \delta} dx^{\gamma} \wedge
dx^{\delta}$.  The corresponding Chern polynomials can be shown to be
closed $2k$-forms, hence defining cohomology classes on the manifold.
}. This equation can be proven using Newton's identitites
\cite{Roe,Frankel}.
Comparing the recursion relation \eq{eq:recursion} with
\eq{schur-recur}, we see that the expressions for Schur polynomials,
$s_k$'s, in terms of polynomials of the traces $Tr X^m$ of the matrix $X$ are identical to
the expressions for the oscillators $\tilde{\sigma}_k^{\dagger}$ in
terms of polynomials of $\beta_m$'s (acting on vacuum). 
In the following we will denote these polynomials of
$\beta_k$'s also as $s_k$'s, which is justified because of the 
equivalence between the $\beta_m$'s and single trace operators, $Tr X^m$. 

The multi-particle states  $\ad_{m_1} \ad_{m_2}...\ket 0$
are, however, different from the states created
by the Schur polynomials. For example, we have 
\bea
(s^\dagger_1)^2 \ket{F_0} &&= (\betad_1)^2 \ket{F_0}
\nn
&& = \half \sqrt{N(N-1)}\sigmad_2\ket 0 + \half\sqrt{N(N+1)}
(\sigmad_1)^2 \ket 0.
\label{multi-schur}   
\eea
We see that the multi-Schur state is in general a linear combination
of multi-particle states of the bosonic oscillators. Moreover,
using the fact that 
\[
s^\dagger_2 \ket{F_0} = \frac{1}2\sqrt{N(N-1)}\sigmad_2 \ket 0
\]
we find, using the definition $||s_i|| \equiv ||s_i^\dagger \ket{F_0}||$,
\[
\frac{\langle F_0 | s_1^2 s^\dagger_2 \ket{F_0}}{||s_1||^2~||s_2||}
= \sqrt{1- 1/N},
\] 
which is an example of the  $O(1)$  interaction among the Schur
polynomial states (see \eq{o-1-schur}). This shows once again that multi-Schur 
states are different from multi-particle states of the bosonic oscillators.

\section{\label{discussion}Summary and discussion}

In this paper we have studied non-perturbative quantum dynamics of
the LLM (half-BPS) fluctuations around AdS$_5 \times$S$^5$ using its
correspondence to the boundary super Yang-Mills theory. We have seen
that a description of the bulk gravitational physics in terms of the
perturbative graviton states breaks down at sufficiently high
energies. This is expected in any theory of gravity. But in the
example studied here, we can go further and identify the new weakly
coupled degrees of freedom in terms of which the bulk physics must be
described at high energies. We have argued that these are the giant
graviton states. A remarkable thing about the LLM sector is that all
the states in this sector, namely gravitons, giants, Schurs, etc. can
be described in terms of the set of $N$ free bosonic oscillators $a_k,
\ad_k$. From this point of view interactions emerge as a result of the 
fact that gravitons, giants, Schurs, etc. are linear combinations of
multi-particle states created by these oscillators.  

An interesting feature of gravity in the LLM sector is that
perturbation theory remains valid until energies of order $N^{1/2}$
are reached. General arguments for validity of perturbative gravity
would have given the relevant (dimensionless) scale to be $R/l_p \sim
N^{1/4}$. Presumably the high degree of supersymmetry in the LLM
sector is responsible for this, but it would be interesting to have an
explicit argument. A related fact \cite{Bala} is that while the size
(in AdS$_5$) of a graviton excitations becomes smaller than the
$10-$dimensional Planck scale for energies larger than $N^{1/2}$, the
size of giant gravitons on $S^5$ becomes larger than Planck scale for
angular momenta larger than $N^{1/2}$. This also seems to suggest that
a meaningful description of physics in the bulk can be constructed in
terms of giant gravitons precisely at and beyond those energies where
one might expect the graviton description to break down.

Another question that has arisen from the present investigation is
about the identification of the boundary states corresponding to
multi-giants. We have seen that at the level of single-particle
states, Schurs in totally antisymmetric represenatations are identical
to the states created by the oscillators $\ad_k$. However, since
muti-Schur states are different from multi-$\ad_k$ states, the former
being interacting while the latter are free, one might ask
which of these correspond to the bulk multi-giant states. Clearly,
this question can only be settled by computations of interactions of
giants among themselves and with gravitons (all staying within the
half-BPS sector) in the bulk string theory or its semiclassical
gravity limit.

\gap5

\noindent{\bf Acknowledgments}

We would like to thank N. Suryanarayana for comments on the manuscript
and for pointing out reference \cite{Okuyama} to us. 


\appendix

\section{\label{a.sec.betas}Beta's and oscillators}

In this Appendix we will collect some useful formulae about the
graviton operators, $\betad_k$. Acting on the fermi vacuum and using
the bosonization formulae, we can write a simple expression for
$\betad_k | F_0 \rangle$ in terms of the bosonic oscillators. We get,
\bea
\beta_m^\dagger~\ket{F_0} = 
&& \sum_{n=2}^N (-1)^{n-1}~\sqrt{\frac{(N+m-n)!}{2^m (N-n)!}}~
\theta(m-n)~{\sigmad_1}^{m-n}~\sigmad_n~\ket{0} \nn
&& ~~~~~~~~~~~~~~~~~~~~~~~~~~~~ 
+\sqrt{\frac{(N+m-1)!}{2^m (N-1)!}}~{\sigmad_1}^m~\ket{0}.
\label{singleparticle}
\eea
Note that the $\sigma_k$'s are related to the oscillator $a_k$'s 
by (\ref{sigmadefs}). Also, $\theta(m)=1$ if $m \geq 0$, otherwise
it vanishes. We list below the first few of these explicitly:
\bea
\betad_1 \ket{F_0} &&= \sqrt {\frac N 2} \sigmad_1\ket 0,
\nn
\betad_2 \ket{F_0} &&= -\half \sqrt{N(N-1)} \sigmad_2\ket 0
+ \half\sqrt{N(N+1)} (\sigmad_1)^2 \ket 0, 
\label{examples}
\eea
etc. The action of multiple $\beta$'s on fermi vacuum can also be
expressed in terms of the bosonic oscillators. For example,
\bea 
(\betad_1)^2 \ket{F_0}=
\half \sqrt{N(N-1)}\sigmad_2\ket 0 + \half\sqrt{N(N+1)}
(\sigmad_1)^2 \ket 0,
\label{examples-many}
\eea
etc. This requires a more general bosonization formula than 
\eq{singleparticle}, which can be easily obtained using the 
discussion in Sec 2. Finally, we note that beyond $m=N$, 
there is no single-particle piece on the right hand side. 
For example,
\bea
\beta_{N+1}^\dagger~\ket{F_0} =
&& \sum_{n=2}^N (-1)^{n-1}~\sqrt{\frac{(N+m-n)!}{2^m (N-n)!}}~
~{\sigmad_1}^{N+1-n}~\sigmad_n~\ket{0} \nn
&& ~~~~~~~~~~~~~~~~~~~~~~~~
+\sqrt{\frac{(2N)!}{2^{N+1} (N-1)!}}~{\sigmad_1}^{N+1}~\ket{0},
\label{fuzzy-gravitons}
\eea
which is a linear combination of multi-particle states, with no 
single-particle component.

\section{Derivation of Eqn.(3.21)}\label{sec:3-point}

We begin by noting the useful formula:
\bea 
\frac{(N+p_1)!}{(N - p_0 -1)!}
&& =  N^{p_0 + p_1 + 1} \exp[\sum_{k=0}^{p_0+ p_1} \ln(1 + \frac{p_1-k}N)]
\nn
&&= N^{p_0 + p_1 + 1}\exp[ \frac{1}{2N}(p_1 - p_0)(p_0+ p_1+1) + 
O(\frac1{N^2})]
\eea
where the first line is exact and the second line is valid for
$p_0/N, p_1/N \ll 1$. The $O(1/N^2)$ term actually evaluates
to $-\frac{1}{6N^2}(p_1^2 + p_0^2 - p_0 p_1 + \frac{p_0 + p_1}{2})(p_0+ p_1+1)$.

Using this approximation repeatedly, it is easy to prove
\bea
\begin{split}
\langle \beta_m \betad_m \rangle & \approx 2T(m)
  \sinh\frac{m(m+1)}{2N}, \\
\langle \beta_m \beta_n \betad_{m+n} \rangle & 
\approx 4T(m+n)
\sinh\left(\frac{m(m+n+1)}{2N}\right)
\sinh\left(\frac{n(m+n+1)}{2N}\right),
\end{split}
\eea
where
\bea
  T(m) \equiv \frac{N^{m+1}}{2^m (m+1)}.
\eea
For $m \sim n$, 
the neglected terms (in the arguments of the $\sinh$)
are of order $m^3/N^2$. Compared to the leading term ($\sim m^2/N$), 
this is down by a factor of $m/N$. These can, therefore,
be neglected if we assume $m/N \ll 1$. We can now use the
two equations above to arrive at \eq{3ptapprox}.

\section{Proof of Eqn.(3.28)}\label{sec:proofrec}

\begin{figure}[htb] \centering
\includegraphics[height=7cm]{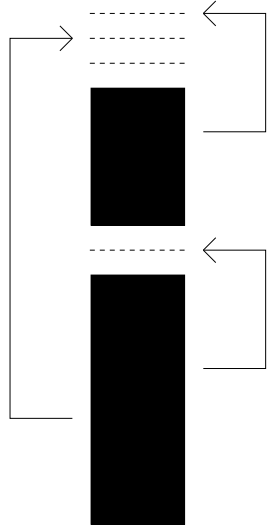} \caption{Forming the state
$\beta_m^{\dagger} \sigma_{k-m}^{\dagger} \ket{0}$.}
\label{fig:betaastate} 
\end{figure}

In this section, we will prove equation (\ref{eq:recursion}).
We will utilize the intuition derived from the bosonization picture.
The fundamental object of interest to us is $\beta_m^{\dagger}
\sigma_{k-m}^{\dagger} \ket{0}$.  First, the bosonic creation operator
$\sigma_{k-m}^{\dagger}$ lifts the $(k-m)$ top fermions by one step,
creating a hole at level $N-k+m$. The action of the graviton operator 
$\beta_m^{\dagger}$, \eq{def-beta}, on this state is to lift a fermion in any one 
of the occupied levels by another $m$ steps, which can happen in three 
qualitatively different ways, as shown in
figure \ref{fig:betaastate}.  
In the first case, indicated by the lower right arrow in
Fig.\ref{fig:betaastate}, one trades a hole for another in the lower
heap of occupied levels. Alternatively, one may place the fermion from
the lower heap (which requires the level $n$ of the fermion
annihilation operator, see \eq{def-beta}, to satisfy $n \leq
(N-k+m-1)$) on top of everything (which requires the level $(m+n)$ of
the fermion creation operator to satisfy $(m+n) \geq (N+1)$). This
case is indicated by lower left arrow in Fig.\ref{fig:betaastate}. In
this case, the topmost ``chunk" consists only of a single
fermion. Finally, if the fermion originated from the top heap, as
indicated by the top right arrow in Fig.\ref{fig:betaastate}, we must
satisfy the conditions $n \ge (N-k+m+1)$ and $(m+n) \geq (N+1)$.

The final state obtained in this way can be described in terms of 
bosonic creation operators acting on the vacuum. We need to be 
careful about relative signs arising from moving the
annihilation operator $\psi_n$ of equation (\ref{def-beta}) down
by $m$ steps as compared to the position of the
$\psi_{n+m}^{\dagger}$, passing through chunks of fermions on its way.
Taking all this into account, the result turns out to be
\begin{equation}\label{eq:bav}
  \begin{split} \beta_m^{\dagger} \sigma_{k-m}^{\dagger} \ket{0} & =
  (-1)^{m-1} C(m,N-k) \sigma_k^{\dagger} \ket{0} \\ & +
  \sum_{n=N-m+1\mbox{ (} m \le k/2 \mbox{)} }^{n=N} (-1)^{N-n} C(m,n)
  \sigma_{k-m}^{\dagger} \sigma_{N-n+1}^{\dagger}
  (\sigma_1^{\dagger})^{n+m-N-1} \ket{0} \\ & + \sum_{n=N-k+m+1\mbox{
  (} m > k/2\mbox{)} }^{n=N} (-1)^{N-n} C(m,n) \sigma_{k-m}^{\dagger}
  \sigma_{N-n+1}^{\dagger} (\sigma_1^{\dagger})^{n+m-N-1} \ket{0} \\ &
  + \sum_{n=N-m+1\mbox{ (} m > k/2\mbox{)} }^{n=N-k+m-1} (-1)^{N-n-1}
  C(m,n) \sigma_{N-n}^{\dagger} \sigma_{k-m+1}^{\dagger}
  (\sigma_1^{\dagger})^{n+m-N-1} \ket{0}, \end{split}
\end{equation}
where the three sums only contribute for $m \le k/2$, $m > k/2$ and $m
> k/2$, respectively, as indicated. Here and in the following
we have assumed that $k$ is even. The odd $k$ case can be handled similarly.

In principle, we should now use the expression (\ref{eq:bav}) to prove
that (\ref{eq:recursion}) holds at a general level $k$. However,
(\ref{eq:recursion}) can actually be split into two parts which cancel
independently of each other. The first part is
\begin{equation}\label{eq:recursionpart1}
  \begin{split}
\sum_{m=1}^{k} (-1)^{k-m} (-1)^{m-1} C(m,N-k) C(k-m,N+m-k) 
\sigma_k^{\dagger} & \ket{0} \\
    + (-1)^k k C(k,N-k) \sigma_k^{\dagger} & \ket{0} = \\
    = (1-1) (-1)^{k-1} k C(k,N-k) \sigma_k^{\dagger} \ket{0} & = 0.
  \end{split}
\end{equation}
where $C(m,N-k) C(k-m,N+m-k) = C(k,N-k) $ was used.
Using (\ref{eq:bav}) and (\ref{eq:recursionpart1}), what then remains
to prove (\ref{eq:recursion}) is that
\begin{equation}\label{eq:recursionpart2}
   \begin{split}
      0 = & \sum_{1 \le m \le k/2} (-1)^{k-m} 
\sum_{n=N-m+1}^{n=N} (-1)^{N-n} \alpha_{mn} \sigma_{k-m}^{\dagger} 
\sigma_{N-n+1}^{\dagger} (\sigma_1^{\dagger})^{n+m-N-1} \ket{0} \\
      + & \sum_{k/2 < m \le k} (-1)^{k-m} \sum_{n=N-k+m+1}^{n=N} 
(-1)^{N-n} \alpha_{mn} \sigma_{k-m}^{\dagger} \sigma_{N-n+1}^{\dagger} 
(\sigma_1^{\dagger})^{n+m-N-1} \ket{0} \\
      + & \sum_{k/2 < m \le k} (-1)^{k-m} \sum_{n=N-m+1}^{n=N-k+m-1} 
(-1)^{N-n-1} \alpha_{mn} \sigma_{N-n}^{\dagger} 
\sigma_{k-m+1}^{\dagger} (\sigma_1^{\dagger})^{n+m-N-1} \ket{0},
    \end{split}
\end{equation}
where
\begin{equation}
  \alpha_{mn} \equiv C(m,n)C(k-m,N+m-k).
\end{equation}

To complete the proof, we proceed as follows. To make summands more
equal, shift $n \rightarrow n-1$ and $m \rightarrow m+1$ in the last
sum over the index $m$.  Redefining
\begin{equation}
  \begin{split}
    n & \equiv N-q+1 \\
    m & \equiv k-p
  \end{split}
\end{equation}
and dividing by $(-1)^k$
then turns (\ref{eq:recursionpart2}) into
\begin{equation}\label{eq:remainstoprove}
  0 = \sum_{p=k/2}^{k-1} \sum_{q=1}^{k-p} \gamma_{pq} \sigma_{pq}
    + \sum_{p=1}^{k/2-1} \sum_{q=1}^{p} \gamma_{pq} \sigma_{pq}
    - \sum_{p=1}^{k/2} \sum_{q=p}^{k-p} \tilde{\gamma}_{pq} \sigma_{pq},
\end{equation}
where $\gamma_{pq}$ really just is shorthand for $\alpha_{mn}$ in the
new indices,
\begin{equation}
\gamma_{pq} \equiv \alpha_{m=N-q+1 \mbox{, } n=k-p} = C(p,N-p) C(k-p,N-q+1).
\end{equation}
Furthermore,
$\tilde{\gamma}_{pq}$ is the analogous coefficient in the third sum over 
the index $m$, in which
the $m \rightarrow m+1$ ,$n \rightarrow n+1$ shifts were performed, i.e.
\begin{equation}
  \begin{split}
    \tilde{\gamma}_{pq} & \equiv C(k-m-1, N+m-k+1) C(m+1,n-1) = \\ 
                        & = C(p-1,N-p+1) C(k-p+1,N-q).
  \end{split}
\end{equation}
The virtue of (\ref{eq:remainstoprove}) is that all the sums are 
now written in terms of
the fundamental variable
\begin{equation}
  \sigma_{pq} \equiv (-1)^{p+q} \sigma_p^{\dagger} \sigma_q^{\dagger} 
\left( \sigma_1^{\dagger} \right)^{k-(p+q)} \ket{0}.
\end{equation}

Changing the order in which the terms in the last sum over the index
$p$ are summed using
\begin{equation}
\sum_{p=1}^{k/2} \sum_{q=p}^{k-p} = \sum_{q=1}^{k/2-1} \sum_{p=1}^q + 
\sum_{q=k/2}^{k-1} \sum_{p=1}^{k-q},
\end{equation}
swapping $p$ for $q$ in that sum, 
and using the facts that $\tilde{\gamma}_{qp} = \gamma_{pq}$ and 
$\sigma_{qp}=\sigma_{pq}$ finally 
turns the equation (\ref{eq:remainstoprove}) that we want to prove into
\begin{equation}
  \begin{split}
  0 & = \sum_{p=k/2}^{k-1} \sum_{q=1}^{k-p} \gamma_{pq} \sigma_{pq}
    + \sum_{p=1}^{k/2-1} \sum_{q=1}^{p} \gamma_{pq} \sigma_{pq} \\
    & - \sum_{p=1}^{k/2-1} \sum_{q=1}^{p} \gamma_{pq} \sigma_{pq}
    - \sum_{p=k/2}^{k-1} \sum_{q=1}^{k-p} \gamma_{pq} \sigma_{pq},
  \end{split}
\end{equation}
which is trivially true, hence concluding the proof of (\ref{eq:recursion}).

     \newcommand{\sbibitem}[1]{\bibitem{#1}}
     
     \end{document}